# Using Multiple Code Representations to Prioritize Static Analysis Warnings

Thanh Trong Vu
*Faculty of Electronics and Telecommunications*
*VNU University of Engineering and Technology*
Hanoi, Vietnam
19020626@vnu.edu.vn

Hieu Dinh Vo
*Faculty of Information Technology*
*VNU University of Engineering and Technology*
Hanoi, Vietnam
hieuvd@vnu.edu.vn

*Abstract*—In order to ensure the quality of software and prevent attacks from hackers on critical systems, static analysis tools are frequently utilized to detect vulnerabilities in the early development phase. However, these tools often report a large number of warnings with a high false-positive rate, which causes many difficulties for developers. In this paper, we introduce VULRG, a novel approach to address this problem. Specifically, VULRG predicts and ranks the warnings based on their likelihoods to be true positives. To predict these likelihoods, VULRG combines two deep learning models CNN and BiGRU to capture the context of each warning in terms of program syntax, control flow, and program dependence. Our experimental results on a real-world dataset of 6,620 warnings show that VULRG's Recall at Top-50% is 90%. This means that using VULRG, 90% of the vulnerabilities can be found by examining only 50% of the warnings. Moreover, at Top-5%, VULRG can improve the state-of-the-art approach by +30% in both Precision and Recall.

*Index Terms*—Vulnerability, static analysis warnings, program dependence graph, control flow, abstract syntax tree

## I. INTRODUCTION

Nowadays, software is applied in almost all of the essential fields in our life. Therefore, issues related to security and software quality are increasingly concerning. To ensure the quality and prevent potential attacks, static analysis (SA) tools are often utilized to detect software vulnerabilities in the early development phase [1, 2]. However, a well-known limitation of these tools is that they report too many false-positive warnings. Contradict to *true positive warnings* or *true positives* (TPs), which are actual vulnerabilities, *false positive warnings* or *false positives* (FPs) are the positions which are non-vulnerable, yet incorrectly reported by SA tools.

A severe consequence of the high number of FPs is reducing the productivity of developers [3, 4]. They have to waste time and effort investigating many safe code segments to find vulnerabilities, yet do not obtain any benefit. Realizing that issue, in recent years, not only the accuracy of the SA tools are significantly improved to reduce FP warnings but also many approaches about post-handling SA warnings are proposed to address this problem [5, 6, 7, 8].

There are multiple studies [5, 6] represent warnings by the fixed sets of the hand-engineered features which are designed based on warning information and/or code metrics. Then, statistical machine learning models are applied to predict whether an SA warning is TP or FP. However, these features are often manually designed by experts and they are specific for several types of warnings. Therefore, it is difficult to extend these approaches to address warnings of different SA tools or to classify other types of warnings.

In addition, with the development and success of deep learning, several recent approaches proposed deep learning models to better learn and capture the patterns associated with TP and FP warnings. This direction has obtained promising results. For instance, Lee et al. [7] built a Convolutional Neural Network (CNN) which exploits the lexical surrounding context of a warning to classify whether this warning is TP or FP. Meanwhile, instead of classifying the warnings into "true" or "false" explicitly, Ngo et al. [8] ranked the warnings according to their probabilities to be TPs. Their approach applied slicing techniques to extract the dependent contexts of the reported statements and train two different Bidirectional Long Short-Term Memory (BiLSTM) models to extract meaningful indications to rank input warnings.

In this paper, we present a novel approach to prioritize warnings produced by SA tools. Specifically, for a set of warnings, VULRG predicts their likelihoods to be TPs based on their contexts about program dependence, syntax, and execution order. After that, these warnings are ranked according to their predicted scores. To represent the warning's context, VULRG not only analyzes its dependencies but also takes into account the syntax and execution order of the program. The reason is that to determine whether the reported statement is vulnerable or not, we need to examine not only that statement but also the statements impact/be impacted by it, to understand its behaviors. Also, the execution order of the statements and their syntax are essential for models to understand the warnings and more accurately capture the patterns associated with TP and FP warnings. In this work, to obtain such information, VULRG uses program dependence graph (PDG) [9] and control flow abstract syntax tree (CF-AST).

Our experimental results on 6,620 warnings of 10 real-world projects show that VULRG ranks 68.2% of vulnerabilities in Top-20%. In addition, for Buffer Overflow (BO) vulnerabilities, VULRG's Recall at Top-50% is 90.9% and this figure for Null Pointer Deference (NPD) vulnerabilities is 99%. This means that, with VULRG, by investigating only 50% of the warnings, developers can find about 91% and 98% of actual

BO and NPD vulnerabilities, respectively. These figures for the state-of-the-art approach, DEFP [8], are 81.68% and 89.74%. Interestingly, at Top-5%, VULRG can improve DEFP by more than 30% in both Precision and Recall.

The rest of the paper is organized as follows: Sec. II represents our approach to prioritize the SA warnings by multiple context representations: PDG and CF-AST. Next, our evaluation setup and experimental results are shown in Sec. III and Sec. IV, respectively. Sec. V discusses the related studies. Finally, our conclusion is summarized in Sec. VI.

## II. RANKING STATIC ANALYSIS WARNINGS

Fig. 1 shows the overview of our approach. In order to precisely capture the patterns associated with TP and FP warnings, VULRG takes into account both semantic and syntactic information of the warnings. First, the context of warnings is represented by program dependence, syntax, and execution order. For program dependence, we apply slicing techniques on PDG to extract statements which impact/be impacted by the reported statement via control/data dependence. The program syntax and execution order of the statements are obtained by analyzing CF-AST. Second, the contexts of the warnings are converted into numeric vectors. Third, these vectors are continuously fed to CNN and BiGRU models. According to the predicted scores to be TPs of the warnings, VULRG outputs a ranked list of the SA warnings.

### A. Context Extraction

*1) Program slice:* To understand the behaviors of the reported statement $s$, we analyze not only $s$ but also the context of $s$. The context of $s$ is represented by *program slice* which is a set of statements impacting/being impacted by $s$ via control/data dependence. Indeed, in a program, a statement cannot execute solely, it often interacts with the other statements in the program to construct the program's behaviors. To capture such important information, we employ PDG [9] and apply slicing techniques [10] to extract all the statements in the program which impact and are impacted by the corresponding reported statement. Specially, we use Joern [11] and conduct both backward and forward interprocedural slicing in the program.

*2) Code gadget:* Besides semantic information (i.e., program slice) of the reported statement $s$, the syntactic information and execution order of $s$, as well as the other statements in the program are also important to reason about vulnerability/non-vulnerability of $s$. In this work, we analyze CF-AST to capture both program syntax and execution order information. Such information is represented in a *code gadget* which is a set of statements with the information about the execution order and the functionality of code elements are encoded. To construct CF-AST, we use PyCParser [12] to build AST and Joern [11] to extract control flow information.

### B. Context Vectorization

Before feeding the extracted contexts into neural network models, these inputs needs to be represent in a format compatible with respective networks. Thus, the goal of this phase is to convert input context from a sequence of lexical code tokens into uniform length vector representations while still preserving the syntactic and semantic information of the original form. The process of vectorizing a code snippet includes 3 steps: (1) Preprocessing (2) Tokenizing (3) Embedding.

*1) Preprocessing:* After extracting relevant context, remaining code statements might still contain comments and special characters (e.g., tab, backslash, end-of-line characters) which might bring noises to the model in the learning stage. To keep only code content that is essential for analyzing the warnings, all comments are removed and special characters are altered by space tokens. In addition, we observe that programmers have variances in coding styles, especially in naming convention for identifiers such as variables and functions. This may cause the neural network models, which are directly trained on raw sources, misled by encoding specific naming characteristics and simply imply them to specific warning labels. To mitigate this issue and enable the models to be able to explore the general vulnerability patterns, VULRG abstracts all the identifiers and constants in the context. For example, a specific function name, variable name, or constant value are replaced by token FUNC, VAR, or LITERAL, respectively.

*2) Tokenizing:* VULRG employs lexical analysis to break down each code statement into a sequence of tokens, including identifiers, constants, keywords, operators, and punctuation marks. For instance, a function-call statement:

```
copy_data(u_in, u_out);
```

is abstracted into

```
FUNC1(VAR1, VAR2);
```

and tokenized into 7 code tokens:

"FUNC1", "(", "VAR1", ",", "VAR2", ")" and ";"

In fact, the number of code tokens in each sequence after being tokenized could be significantly different. Thus, VULRG defines a fixed length, $l$, and pad/truncate all the input sequences to fit this length. Particularly, for sequences having lengths smaller than $l$, VULRG pads one or more special tokens (<pad>) at the end of these sequences. Meanwhile, for the sequences whose lengths are greater than $l$, they will be truncated gradually from both sides to reach the length $l$. Note that, the value of $l$ is carefully selected via multiple experiments.

*3) Embedding:* To make the program syntactically correct and precisely convey semantic information of source code [13], code tokens have to appear together in a certain order. For this purpose, in the embedding step, VULRG employs unsupervised Word2vec model [14], which has demonstrated its effectiveness in multiple code mining tasks [15], to encode relationships between neighboring code tokens. In particular, after being trained on the whole code corpus, Word2vec is able to transform a code token into a $k$-dimensional vector. Hence, the embedding of a code context, which corresponds to a sequence of $l$ code tokens, is a $l \times k$ matrix.

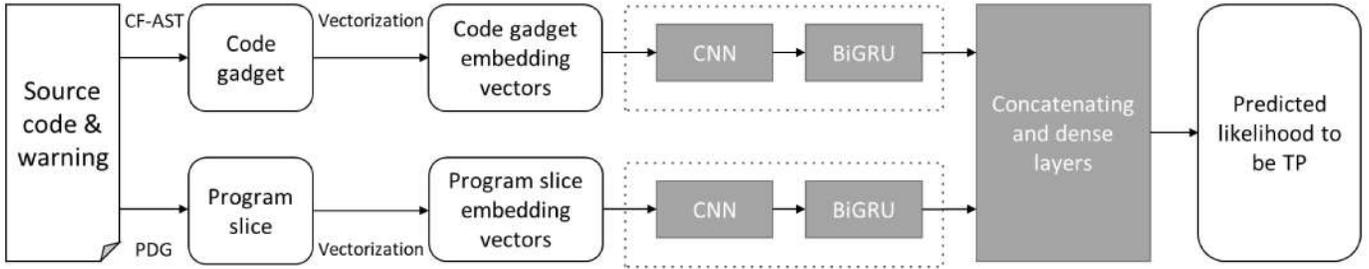

Fig. 1: VULRG's approach overview

## C. Representation Learning and Warning Ranking

In this work, we employ learning-based models to learn and represent the extracted contexts (i.e., program slice and code gadget) to apprehend their long-term dependencies. The reason is that, although unsupervised embedding models (e.g., Word2vec) could interpret syntactic and semantic information to a certain degree, its limitation is that it only learns correlations between tokens in specific window sizes, and could cause scalability issues in capturing such information on a larger scope if code tokens lying in different code statements.

In practice, to learn the representation of an arbitrary text sequence, Gated Recurrent Unit (GRU) and Long short-term memory (LSTM) are the most popular choices, which could obtain both the contextual information and the order of tokens. Meanwhile, despite the similar performance of GRU and LSTM in almost cases, GRU brings the advantage of more computational efficiency. However, GRU is primarily designed to pass input as one-way time steps in sequential order, in the meantime, while applying for code, the occurrence of a code token is often related to the previous and/or the next tokens. Thus, to extract dependencies in both forward and backward order, VULRG constructs stacked GRU networks following the bidirectional architecture (so-called BiGRU).

Furthermore, although the program slice and code gadget often contain relevant information to help analyze SA warnings, they might still contain noises that cause a negative impact on representation models. Even when BiGRU model holistically observes input sequence, it might still be biased by these noises. Thus, to assist BiGRU to focus on important features from the input context, CNN is employed to produce intermediate representative feature maps.

In the final layer, VULRG applies two Global Max Pooling (GMP) layers following each of BiGRU networks and then concatenated their outputs into a unified one to form the final representation (Fig. 1). This representation is later fed into multiple Fully Connected Neural Networks (FCNN) with a softmax activation function to predict the likelihood to be TP or FP of the current warning.

## III. EMPIRICAL METHODOLOGY

In order to evaluate VULRG, we seek to answer the following research questions:

TABLE I: Dataset overview

| No. | Project | Buffer Overflow | | | Null Pointer Dereference | | |
|---|---|---|---|---|---|---|---|
| | | #W | #TP | #FP | #W | #TP | #FP |
| 1 | Asterisk | 2,049 | 63 | 1,986 | 133 | 0 | 133 |
| 2 | FFmpeg | 1,139 | 387 | 752 | 105 | 37 | 68 |
| 3 | Qemu | 882 | 396 | 486 | 72 | 39 | 33 |
| 4 | OpenSSL | 595 | 53 | 542 | 9 | 2 | 7 |
| 5 | Xen | 388 | 15 | 373 | 23 | 6 | 17 |
| 6 | VLC | 288 | 20 | 268 | 16 | 2 | 14 |
| 7 | Httpd | 250 | 45 | 205 | 17 | 0 | 17 |
| 8 | Pidgin | 250 | 13 | 237 | 242 | 0 | 242 |
| 9 | LibPNG | 83 | 9 | 74 | 2 | 0 | 2 |
| 10 | LibTIFF | 74 | 9 | 65 | 3 | 3 | 0 |
| # | Total | 5,998 | 1,010 | 4,988 | 622 | 89 | 533 |

#W, #TP and #FP are total warnings, true positives and false positives.

- **RQ1:** How accurate is VULRG in ranking SA warnings? and how is it compared to the state-of-the-art approach [8]?
- **RQ2:** How do the deep learning models impact the performance of VULRG?
- **RQ3:** How do the context lengths impact the performance of VULRG?

## A. Dataset

There are several synthetic datasets containing a large number of vulnerabilities which can be used to evaluate our approach, such as Juliet [16] and SARD [17]. However, the previous study [18] showed that artificial vulnerabilities are too simple and they cannot be used to estimate the performance of the machine learning models on a real-world dataset. In this work, we aim to evaluate the performance of VULRG on a real-world dataset, so we reuse the benchmark proposed by Ngo et al. [8]. In total, there are 6,620 warnings in 10 popular open-source C/C++ programs. These warnings are categorized into two vulnerability types: *Buffer Overflow (BO)* and *Null Pointer Dereference (NPD)*. The overview of this dataset is shown in Table I.

## B. Empirical procedure

**RQ1.** We compare the performance of VULRG and the state-of-the-art approach, DEFP [8], in ranking static analysis warnings. For the sake of comparison, we adapt their experimental settings: within-project setting and combined-project setting, which are also widely used in related studies [19, 20, 21]. Specifically, for *within-project setting*, warnings from the same project are split into training and testing sets. Meanwhile, for *combined-project setting*, warnings from all 10 projects are shuffled and then split into training and testing sets. Note that, the *within-project setting* is adapted for three projects: Qemu, FFmpeg, and Asterisk, which contain the largest number of BO warnings. This setting can not be applied in the other projects because of their small number of warnings which cannot be used to train and evaluate a machine learning model.

**RQ2.** We study the impact of the deep learning models on the performance of VULRG. Specially, we compare the performance of VULRG in three settings: (1) using only CNN models, (2) using only BiGRU models, (3) combining CNN models and BiGRU models.

**RQ3.** We study the impact of context length on the performance of VULRG by gradually changing the limit lengths (the value of $l$ which is defined in. Sec. II-B2) of code gadgets and program slices. Firstly, we fix the program slice limit length to 300 tokens, and change the code gadget limit length from 300 to 1200 tokens. Secondly, we fix the code gadget limit length to 600 tokens, and change the program slice limit length from 300 to 800 tokens.

### C. Metrics

In order to evaluate VULRG and compare its performance with DeFP, we employed Top-$k$% Precision (P@K) and Top-$k$% Recall (R@K). P@K and R@K are calculated using the following formulas:

$$P@K = \frac{|\{Actual\ TPs\} \cap \{Predicted\ TPs\}@K|}{|\{Predicted\ TPs\}@K|}$$

$$R@K = \frac{|\{Actual\ TPs\} \cap \{Predicted\ TPs\}@K|}{|\{Actual\ TPs\}|}$$

in which, $\{Actual\ TPs\}$ is the set of actual TP warnings, $\{Predicted\ TPs\}@K$ is the list of Top-$k$% of warnings ranked first by the model.

### D. Experimental setup

In our experiment, the CNN and BiGRU models are implemented using Keras together with TensorFlow backend (version 2.8.0). VULRG used the the tokenizer built upon NLTK library (version 3.6.2) and *Word2vec* embedding model in the gensim package (version 3.6.0). Our experiments were trained and evaluated using Google Colab, Ubuntu 18.04 with an NVIDIA Tesla T4 GPU.

To evaluate the neural network models, we adopt 5-fold cross-validation. In VULRG, the embedding size is set to 64. The other configurations such as the dropout, batch size, and number of epochs are set similarly to DEFP. Specifically, these figures are 0.1, 64, and 60, respectively. Also, the minibatch stochastic gradient descent ADAMAX optimizer is selected with a learning rate of 0.002.

## IV. EVALUATION

### A. Accuracy analysis (RQ1)

Table II shows the performance of VULRG and DEFP in the within-project setting and combined-project setting. By investigating information related to program dependence, execution order, and program syntax, VULRG can capture the patterns of the warnings and effectively predict their likelihood to be TPs. Overall, VULRG obtained 89.02% and 26.40% in Top-5% Precision and Recall. Moreover, at Top-50%, its Precision and Recall are 30.6% and 90.9%. This means that VULRG can help to save a large amount of investigation time and effort, developers can find up to 90% of the actual vulnerabilities by examining only 50% of the warnings.

In addition, VULRG can rank the warnings more correctly than the state-of-the-art approach, DEFP. Especially, at Top-5%, VULRG can improve the performance of DEFP by +30% in Precision and Recall. Also, the improvement at Top-50% is about 12%. This result demonstrates that for the same number of ranked warnings, developers can find more vulnerabilities by investing ranked list of VULRG compared to that of DEFP.

Indeed, both VULRG and DEFP employ dependence information of the reported statements to understand the warnings. Furthermore, VULRG also takes into account the syntax information (AST) and the execution order of the statements (control flow), meanwhile DEFP neglects this information. By using additional information, VULRG is able to interpret surrounding context to have more evidences to better predict whether the reported statements are vulnerable or not.

### B. Impact of the deep learning models (RQ2)

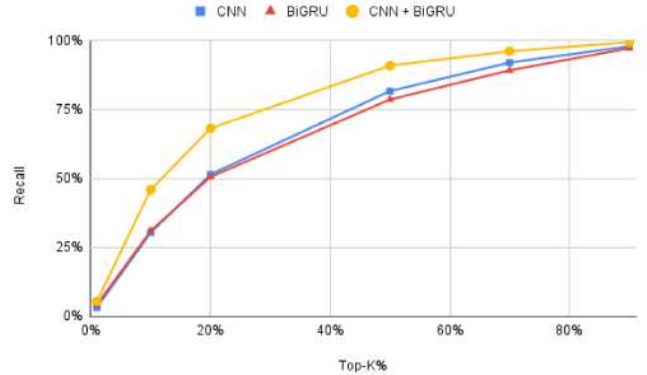

Fig. 2: Impact of models on the performance of VULRG

Fig. 2 shows the performance of VULRG when enabling only CNN models, BiGRU models, and combining them. Overall, VULRG obtained the best performance when both CNN models and BiGRU models are combined. Specifically, using only CNN models, VULRG's Top-50% Recall is 81.7%. This figure when enabling only BiGRU models is 78.62%. Meanwhile, if the features of the embedding vectors of the warnings are captured by CNN models, and then fed to BiGRU models, the performance of VULRG is improved 10%.

TABLE II: Performance of VULRG and DEFP [8] in ranking SA warnings

| WN | Project | Method | # TP warnings found in top-k% warnings | | | | | | | | | |
|---|---|---|---|---|---|---|---|---|---|---|---|---|
| | | | Top-5% | | Top-10% | | Top-20% | | Top-50% | | Top-60% | |
| | | | Precision | Recall | Precision | Recall | Precision | Recall | Precision | Recall | Precision | Recall |
| BO | Qemu | DEFP | 82.50% | 8.74% | 62.5% | 13.26% | 56.38% | 24.62% | 50.74% | 55.04% | 50.20% | 65.08% |
| | | VULRG | 93.34% | 10.62% | 89.98% | 20.46% | 75.42% | 33.34% | 57.5% | 63.88% | 53.78% | 71.96% |
| | FFmpeg | DEFP | 82.50% | 8.74% | 62.50% | 13.26% | 56.38% | 24.62% | 50.74% | 55.04% | 50.20% | 65.08% |
| | | VULRG | 82.50% | 8.74% | 85.00% | 17.98% | 78.76% | 34.40% | 60.72% | 65.90% | 57.34% | 74.36% |
| | Asterisk | DEFP | 31.00% | 48.98% | 17.06% | 55.36% | 10.00% | 64.98% | 4.90% | 79.34% | 4.50% | 87.04% |
| | | VULRG | 51.00% | 81.00% | 25.86% | 84.22% | 13.40% | 87.30% | 5.68% | 92.04% | 4.82% | 93.72% |
| | COMBINED | DEFP | 69.00% | 20.50% | 53.48% | 31.78% | 41.60% | 49.40% | 26.98% | 80.12% | 24.32% | 86.62% |
| | | VULRG | 89.02% | 26.40% | 77.34% | 45.92% | 57.42% | 68.20% | 30.60% | 90.90% | 26.38% | 93.96% |
| NPD | COMBINED | DEFP | 73.34% | 24.62% | 66.66% | 44.90% | 46.40% | 65.24% | 26.44% | 92.14% | 22.58% | 94.36% |
| | | VULRG | 100% | 33.70% | 98.34% | 66.30% | 68.00% | 95.42% | 28.36% | 98.82% | 23.64% | 98.82% |

TABLE III: Impact of code gadget length on VULRG's performance

| Length of code gadget | 300 | 400 | 500 | 600 | 900 | 1200 |
|---|---|---|---|---|---|---|
| Precision | 52.26% | 53.72% | 54.56% | 55.42% | 57.42% | 56.24% |
| Recall | 62.10% | 63.86% | 64.86% | 65.86% | 68.20% | 66.82% |

TABLE IV: Impact of program slice length on VULRG's performance

| Length of program slice | 300 | 400 | 500 | 600 | 700 | 800 |
|---|---|---|---|---|---|---|
| Precision | 56.26% | 55.58% | 55.32% | 57.42% | 57.90% | 56.52% |
| Recall | 66.82% | 66.02% | 65.74% | 68.20% | 68.80% | 67.14% |

These figures prove that the combination of CNN and BiGRU not only helps capture longer contextual dependencies but also guides the models to focus on meaningful features that contribute to the TP/FP prediction.

*C. Impact of context lengths (RQ3)*

Table III and Table IV shows Top-20% Precision and Recall of VULRG with different limit lengths of code gadget and program slice. As seen, the performance of VULRG is slightly affected by the length of code gadgets and program slices. In general, with a longer context, VULRG has more information to understand the warning and obtained better performance. Specifically, VULRG obtains the best performance when the code gadget is limited to 900 tokens and the program slice is limited to 700 tokens. However, if the contexts are too long, they may include noises and negatively impact VULRG's accuracy. When the code gadget length is set to 1200 tokens and the program slice is set to 800 tokens, the performance of VULRG declines about 2%.

V. RELATED WORK

To eliminate FPs, there are various studies [22, 23, 24, 25] used sophisticated verification techniques such as model checking, symbolic execution, deductive verification, etc. In general, these studies handle the problem by proving the violation and/or non-violation of the source code associated with the warnings. However, these approaches are complicated and non-scalable, so it is very difficult to apply in practice.

In recent decades, machine learning models are widely applied to address SA warnings by explicitly classifying them into FPs or TPs groups, or ranking them according to a specific order. There are two main directions in applying machine learning models in this field. Firstly, a warning is represented by manually defined features and then classified by statistical machine learning models [5, 6, 26]. Secondly, deep learning models are built to capture the patterns associated with the warnings, and then these features are used to classify/rank the warnings [7, 8].

Similar to the approach of DEFP [8], we also address the problem of FP warnings by leveraging deep learning models to rank the warnings based on their likelihoods to be vulnerabilities. However, the main differences in our approach compared to DEFP are the analyzing contexts and the employed models. Specifically, to capture the warning patterns, DEFP analyzing the reported statements and their dependencies. In VULRG, not only such information but also the syntax and the execution order of the statements (CF-AST) are considered, which helps VULRG have more comprehensive information to understand the warning. Furthermore, instead of employing only BiLSTM models to represent a code sequence like DEFP, we combine CNN and BiGRU models. By this combination, CNN can help to reduce noise by producing intermediate representative feature maps and BiGRU can represent the contextual information and the order of tokens. This helps VULRG better capture the TP and FP patterns.

VI. CONCLUSION

In practice, static analysis tools are frequently used to detect potential vulnerabilities in the early development phase. However, these tools often report a large number of warnings with a high false-positive rate which reduces the productivity of developers. To address this problem, we introduce VULRG, a novel approach which combines two deep learning models CNN and BiGRU to rank the warning based on their predicted likelihoods to be true positive. In VULRG, to comprehensively understand the warning and precisely rank it, the context of each warning is examined in three aspects: program syntax, control flow, and program dependence. Our experimental results on a real-world dataset of 6,620 warnings show that VULRG's Recall at Top-50% is 90.9%. This means that using VULRG, 90% of the vulnerabilities can be found by examining only 50% warnings. Moreover, at Top-5%, VULRG can improve the state-of-the-art approach by more than 30% in both Precision and Recall.